\documentclass[journal]{IEEEtran}
\usepackage{cite}

\usepackage[pdftex]{graphicx}
\usepackage[caption=false,font=footnotesize]{subfig}
\usepackage{mcite}
\usepackage{url}

\usepackage{color}

\begin{document}

\title{Optimal Load and Stiffness for Displacement-Constrained Vibration Energy Harvesters}

\author{Einar~Halvorsen,~\IEEEmembership{Member,~IEEE,}
\thanks{E. Halvorsen is with the Department
of Micro- and Nanosystem Technology, University College of Southeast Norway, Campus Vestfold, Raveien 215, N-3184 Borre, Norway, e-mail: Einar.Halvorsen@hbv.no.}% 
\thanks{This work was supported by the Research Council of Norway under Grant no. 229716.}%
}

\maketitle

\begin{abstract}
  The power electronic interface to a vibration energy harvester not only provides ac-dc conversion, but can also set the electrical damping to maximize output power under displacement-constrained operation. This is commonly exploited for linear two-port harvesters by synchronous switching to realize a Coulomb-damped resonant generator, but has not been fully explored when the harvester is asynchronously switched to emulate a resistive load. In order to understand the potential of such an approach, the optimal values of load resistance and other control parameters need to be known. In this paper we  determine analytically the optimal load and stiffness of a harmonically driven two-port harvester with displacement constraints. For weak-coupling devices, we do not find any benefit of load and stiffness adjustment beyond maintaining a saturated power level. For strong coupling we find that the power can be optimized to agree with the velocity damped generator beyond the first critical force for displacement-constrained operation.  This can be sustained up to a second critical force, determined by a resonator figure-of-merit, at which the power ultimately levels out.

\end{abstract}


\section{Introduction}

\IEEEPARstart{V}{ibration} energy harvesters are of great interest as power sources for wireless sensors \cite{Mitcheson2008}. In addition to the harvester, its power electronic interface is a central system component. Even though the electronic interface's main purpose is ac-dc conversion, it can also serve other purposes such as effectively increasing the electromechanical coupling \cite{Guyomar2005} or setting the electrical damping to maximize output power under different excitations strengths\cite{Mitcheson2004b}. With a suitable design of the energy harvester, it is even possible to use the electronics to adjust the stiffness to adapt the device to different ambient vibration frequencies \cite{Eichhorn2011}. 

The linear two-port model is perhaps the most widely used model for vibration energy harvesters. Under the right circumstances, it can describe harvesters of either of the main transduction types, that is harvesters with electrostatic \cite{Sterken2003b, Tvedt2010}, piezoelectric \cite{Roundy2004, Goldschmidtboeing2011} or electromagnetic \cite{Amirtharajah1998, Maurath2012} transduction. A linear two-port harvester with a synchronous switching power electronic interface such as the synchronous electric charge extraction (SECE) \cite{Lefeuvre2005}, ideal synchronized switch harvesting on inductor (SSHI) \cite{Guyomar2005} or synchronous pre-biasing \cite{Dicken2012} interface can realize a Coulomb-damped resonant generator (CDRG) \cite{Miller2012}. Hence, these interfaces make it possible to adjust the electric damping to maximize output power under displacement-constrained operation as described for the CDRG in \cite{Mitcheson2004b}. 

Asynchronously switched electronic interfaces, such as buck-boost converters, can appear as an effective load resistance that is determined by the duty cycle of the switching circuit \cite{Dhulst2006,Lefeuvre2007,Dhulst2010}. This circuit alternative performs ac-dc conversion and simultaneously makes possible control of electrical damping through variation of a load resistance. It can be modelled by the much studied, harmonically driven, linear two-port harvester with a resistive load \cite{Amirtharajah1998,duToit2005,Renno2009}. The optimal drive frequency and load resistance of this model are known on closed-form \cite{Renno2009}. 

The resistively loaded two-port model has much in common with a second order model \cite{Williams1996}, the velocity damped resonant generator (VDRG) \cite{Mitcheson2004b}. It is equivalent to the VDRG for short electrical time scale \cite{Amirtharajah1998} and has the same optimal power for strong coupling.
Since the VDRG operating under a displacement-constraint can perform close to rigorous upper bounds on power \cite{Halvorsen2013c}, it is of interest to understand also how an optimal linear two-port model performs under this constraint. While such an analysis has not yet been carried out, maximizing the efficiency, which amounts to maximizing the electrical damping, has shown a clear performance benefit with displacement-constrained operation \cite{Renaud2012}.  

In this paper we  determine analytically the optimum load and stiffness of a harmonically driven linear two-port harvester with displacement constraints. We choose to consider the optimal stiffness instead of the optimal drive frequency which would seem to be the more common point of view. This choice is motivated by the drive frequency normally not being subject to control in an application, while the (anti-)resonant frequency of the device is controllable through the stiffness by design and by electronic means as mentioned in the first paragraph. Since the displacement constraint is enforced by the load and stiffness tuning only, one might be concerned if there are detrimental effects of impacts on end-stops in a real device. To this we take the position that impacts can be avoided by taking a small margin on the electrically controlled displacement limit. 
The mathematical formulation of the problem  treat all three generator types on an equal footing.       

\section{The energy harvesting system} 

\begin{figure}[!t]
\centering
\includegraphics[]{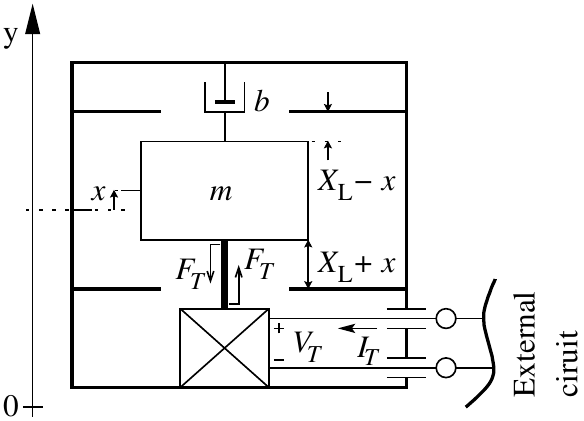}
\caption{Generic energy harvester model.}
\label{fig:twoportgeneric}
\end{figure}

There is a wide range of situations where the dynamic of a  vibration energy harvester is adequately described using only one mechanical coordinate. If the harvester in addition has a single electrical port, we can consider it a two-port device. If we further restrict to linear devices, the mechanical part of the device can be characterized by a mass, a damping and  a stiffness related to the coordinate, and with a linear coupling to the electrical part. The coupling between the electrical and mechanical domain is then conveniently modelled as a linear two-port transducer. Such a model is depicted in Fig. \ref{fig:twoportgeneric} where an inertial mass $m$ is connected to the mechanical port of a transducer and subject to a parasitic damping with constant $b$. When the frame is accelerated,  there will be a relative displacement $x$ between the frame and the mass, and the latter acts on the transducer with a force $F_\mathrm{T}$. The voltage across the terminals of the electric port is $V_\mathrm{T}$ and the current into the positive terminal is $I_\mathrm{T}$.

In a notation adapted from \cite{Tilmans1996,*Tilmans1996errata}, the linear two-port equations for an electrostatic or piezoelectric transducer can be written 
\begin{eqnarray}
F_\mathrm{T} & = & K_1 x + \frac{\Gamma}{C_0} q, \label{eq:espetrdF1}\\
V_\mathrm{T} & = & \frac{\Gamma}{C_0} x +  \frac{1}{C_0}q \label{eq:espetrdV1}
\end{eqnarray}
where $K_1$ is the open  circuit stiffness, $C_0$ is the clamped capacitance and $\Gamma$ is the tranduction factor. The quantity $q$ is the charge, or the deviation in charge from a reference state, on the positive electrical terminal. Note that there is an arbitrariness in the choice of independent and dependent variables. For example, in the piezoelectric litterature it is more common to write these equations 
\begin{eqnarray}
F_\mathrm{T} & = & K_0 x + \Gamma V_\mathrm{T}, \label{eq:espetrdF2} \\
q & = & -\Gamma x +  C_0 V_\mathrm{T} \label{eq:espetrdV2}
\end{eqnarray}
where we have introduced the short-circuit stiffness $K_0=K_1-\Gamma^2/C_0$. 
Other formulations such as the one with $F_\mathrm{T}$ and $V_\mathrm{T}$ as independent variables are also in use \cite{Goldschmidtboeing2011}. Which formulation to choose is entirely a matter of preference, not of physics.
The squared electromechanical coupling factor \cite{Ikeda1996} is often used to characterize the coupling and is given  by 
\begin{equation}
k^2 =\frac{\Gamma^2}{K_1C_0}.
\label{eq:couplingespe}
\end{equation}
 
\begin{figure}[!t]
\centering
\subfloat[Electrostatic or piezoelectric]{%
\includegraphics[height=2cm]{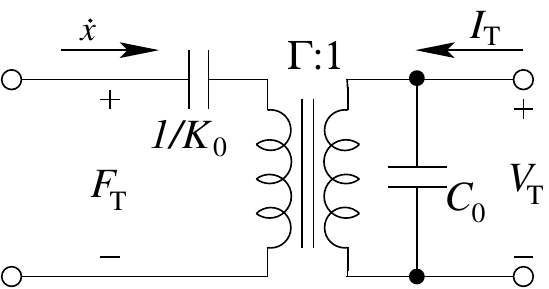}%
\label{fig:espetrd}}\hspace{1em}
\subfloat[Electromagnetic]{%
\includegraphics[height=2cm]{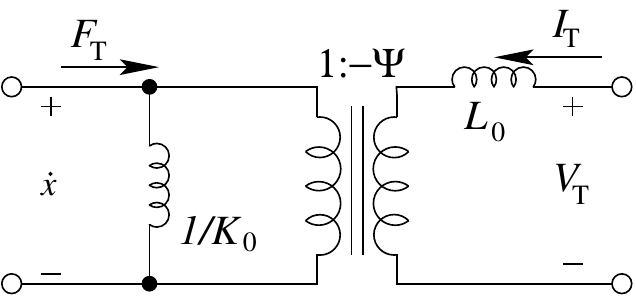}%
\label{fig:emtrd}}
\caption{Equivalent circuits for two-port transducers.}
\label{fig:eqcirctrd}
\end{figure}

In Fig. \ref{fig:espetrd} the transformer and the two capacitors constitute an equivalent circuit for a  lossless two-port transducer with constitutive equations (\ref{eq:espetrdF1},\ref{eq:espetrdV1}) or (\ref{eq:espetrdF2},\ref{eq:espetrdV2}). In this circuit the mechanical domain is represented in the $e-V$ convention where the force is an across variable and the velocity is a through variable \cite{Senturia2001}. 

The constitutive equations of the linearized electromagnetic transducer can be  
written \cite{Tilmans1996,*Tilmans1996errata} 
\begin{eqnarray}
F_\mathrm{T} & = & K_1 x + \frac{\Psi}{L_0} \lambda, \label{eq:emtrdF1}\\
I_\mathrm{T} & = & \frac{\Psi}{L_0} x +  \frac{1}{L_0}\lambda \label{eq:emtrdI1}
\end{eqnarray}
where $\Psi$ is the transduction factor, $L_0$ is the clamped inductance and $\lambda$ is the flux linkage or the deviation in flux linkage from the reference state. In this case, the stiffnes $K_1$ must be interpreted as the short-circuit stiffness. There is an open-circuit stiffness given by  $K_0=K_1-\Psi^2/L_0$ and the electromechanical coupling factor is  
\begin{equation}
k^2 =\frac{\Psi^2}{K_1L_0}.
\label{eq:couplingem}
\end{equation}

Since the structure of (\ref{eq:espetrdF1},\ref{eq:espetrdV1}) and (\ref{eq:emtrdF1},\ref{eq:emtrdI1}) are the same, with flux linkage and charge playing the same role, the topology of Fig. \ref{fig:espetrd} could be used as an equivalent circuit also for the electromagnetic transducer. However, due to the interchange of the roles of current and charge, it is more convenient to choose an alternative such as the circuit in Fig. \ref{fig:emtrd}. This circuit is made in the $f-V$ convention where the velocity is treated as an across variable and the forces are through variables.   

\begin{figure}[!t]
\centering
\subfloat[Electrostatic or piezoelectric generator]{%
\includegraphics[width=2.5in]{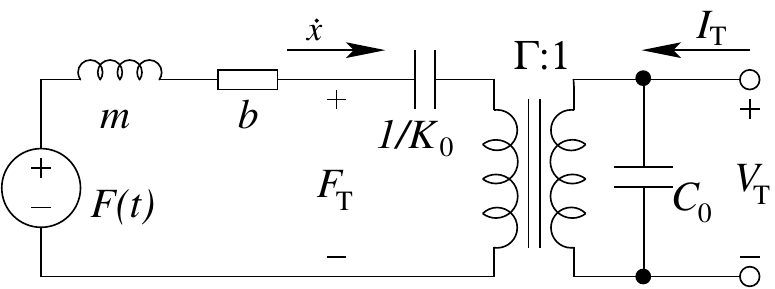}%
\label{fig:espe}}\\
\subfloat[Electromagnetic generator]{%
\includegraphics[width=2.5in]{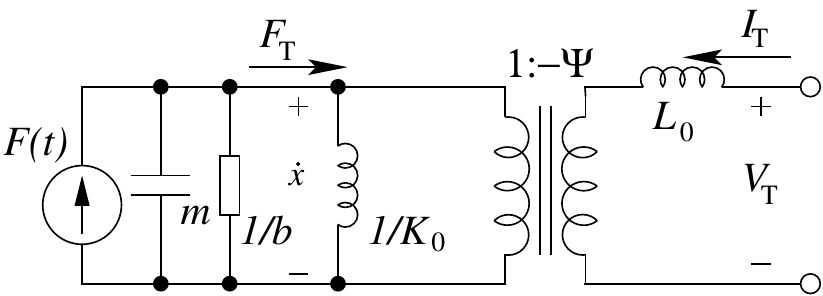}%
\label{fig:em}}
\caption{Equivalent circuits for vibration energy harvesters.}
\label{fig:eqcirc}
\end{figure}

Augmenting the transducer equivalent circuits with mass, damping and a force, we obtain the conventional equivalent circuit representations of linear two-port harvesters in Fig. \ref{fig:eqcirc}. Despite the apparent differences, the two topologies are connected  by a duality relationship which results in similar mathematical equations. 
With an appropriate parametrization, it is possible to express the output power of all the three different architectures on the same form.   

We have denoted the highest mechanical stiffness by $K_1$ and the lowest by $K_0$ for all three transduction types. The corresponding undamped angular resonant frequencies  are then $\omega_1=\sqrt{K_1/m}$  and $\omega_0=\sqrt{K_0/m}$, and the Q-factors are $Q_0=m\omega_0/b$ and  $Q_1=m\omega_1/b$.
The stiffness constants are related by $K_0=(1-k^2)K_1$ with the respective $k^2$ for the two cases. Note that $0\le k^2\le1$. 
Some authors prefer to quantify coupling instead through the coupling factor $k_\mathrm{e}^2=k^2/(1-k^2)\in[0,\infty)$ which is sometimes referred to as the expedient coupling factor \cite{Tadmor2003} or the generalized electromechanical coupling factor \cite{Renaud2012}.  

With a time harmonic drive force $F(t)=F_0\cos\omega t$ of angular frequency $\omega$ and a resistance $R$ connected to the electrical ports in Fig. \ref{fig:eqcirc}, the output power for both  architectures can be written 
\begin{equation}
P =  \frac{1}{2}\Delta K \frac{\omega^2\tau}{1+(\omega \tau)^2} |X_0|^2
\label{eq:power}
\end{equation}
where the complex displacement amplitude is 
\begin{equation}
X_0  = \frac{F_0}{K_1-m\omega^2+i\omega b -\Delta K/(1+i\omega \tau)}
\label{eq:X0}
\end{equation}
and either 
\begin{equation}
\tau = RC_0\,\,\mbox{and}\,\, \Delta K = \Gamma^2/C_0,
\label{eq:taues}
\end{equation}
or
\begin{equation}
\tau = L_0/R\,\,\mbox{and}\,\, \Delta K = \Psi^2/L_0,
\label{eq:tauem}
\end{equation}
depending on the type of generator. By this formulation we have buried the differences between the generator types in the parameter definitions so that a single analysis can cover all three types.

\section{Optimum load and stiffness}

We now maximize the power (\ref{eq:power}) with respect to the stiffness $K_1$ and electrical time constant $\tau$ under the constraint  that $|X_0| \le X_\mathrm{L}$ for a displacement limit $X_\mathrm{L}$. We first treat the case of weak excitations such that the displacement limit is not reached. The stationary points of the power are then determined by $\partial P/\partial K_1=0$ and $\partial P/\partial\tau=0$. The former equation reduces to   
\begin{equation}
K_1 = m\omega^2 + \frac{\Delta K}{1+\omega^2\tau^2}
\label{eq:stiffness}
\end{equation}
which can be used to simplify the latter equation to 
\begin{equation}
(1-\omega^2\tau^2)[b(1+\omega^2\tau^2)-\Delta K\tau]=0.
\label{eq:dPdt}
\end{equation}
From the first factor, one solution is $\omega\tau=1$ leading to $K_1=m\omega^2+\Delta K/2$ and 
\begin{equation}
P = \frac{\Delta K/b\omega}{(2+\Delta K/b\omega)^2}\frac{F_0^2}{b}
= \frac{M}{(2+M)^2}\frac{F_0^2}{b}
\label{eq:pow1}
\end{equation}
where we defined the parameter 
\begin{equation}
M = \frac{\Delta K}{b\omega}.
\end{equation}
We note that the power expression (\ref{eq:pow1}) has a maximum at $M=2$ and that this maximum is $P=F_0^2/8b$ which is the largest possible power these generators can provide. 

The other factor in (\ref{eq:dPdt}) leads to the solution 
\begin{eqnarray}
\omega\tau & = & M/2\pm\sqrt{(M/2)^2-1}, \label{eq:loadhc1}\\
K_1 & = & m\omega^2 + \Delta K \frac{M\mp\sqrt{M^2-4}}{2M}.
\end{eqnarray}
This solution is only real for $M\ge 2$ and always gives output power $P=F_0^2/8b$. The previous solution with lower power applies for $M\le 2$. 

The parameter $M$ is a resonator figure of merit \cite{Vittoz2010}. From (\ref{eq:couplingespe}) and (\ref{eq:couplingem}), we have that $\Delta K = k^2K_1$ and we can write $M=k^2Q_1\omega_1/\omega$ or $M=k_\mathrm{e}^2Q_0\omega_0/\omega$. For moderate coupling $\omega_1\approx\omega_0$ and usually $\omega\approx\omega_{0/1}$, so M is approximately $k_\mathrm{e}^2Q_0$ or  $k^2Q_1$ which have been commonly used  to distinguish the low-coupling from the high-coupling regime.

The displacement amplitude at the optimal stiffness (\ref{eq:stiffness}) is given by  
\begin{equation}
\left.|X_0|^2\right|_{\mathrm{optimal}\;K_1} =
\frac{F_0^2}{\omega^2b^2\left[1+M\omega\tau/(1+\omega^2\tau^2) \right]^2}.
\label{eq:dispopt}
\end{equation}
Inserting (\ref{eq:loadhc1}), we can find the critical force $F_\mathrm{c}$ that makes $X_0=X_\mathrm{L}$ for the optimized strong-coupling case $M\ge 2$ to be
\begin{equation}
F_\mathrm{c}=2b\omega X_\mathrm{L}
\label{eq:fcritical}
\end{equation} 
and with a corresponding output power
\begin{equation}
P_\mathrm{c} = \frac{F_\mathrm{c}^2}{8b}=\frac{1}{2}b\omega^2X_\mathrm{L}^2.
\end{equation}

If we follow the same procedure with $\omega\tau=1$ in (\ref{eq:dispopt}),
we find the critical force in the weak-coupling case $M\le 2$ to be 
\begin{equation}
\frac{1}{2}\left(1+\frac{1}{2}M\right)F_\mathrm{c}
\label{eq:fcritical2}
\end{equation}
which is smaller than (\ref{eq:fcritical}) due to the smaller electrical damping.  

We now consider forcing that is stronger than the relevant critical force, (\ref{eq:fcritical}) or  (\ref{eq:fcritical2}). In this case we need to find the optimal values of stiffness and electrical time scale under the constraint $|X_0|=X_\mathrm{L}$ which we can enforce with a Lagrange multiplier $\lambda$. That is, we look for the optimum among the values of $K_1$, $\tau$ and $\lambda$ that make \begin{equation}
L= \frac{1}{2}\Delta K \frac{\omega^2\tau}{1+(\omega \tau)^2} |X_0|^2 
- \lambda (|X_0|^2-X_\mathrm{L}^2) 
\end{equation}  
stationary. Differentiating and setting equal to zero, we can write the resulting three equations as 
\begin{eqnarray}
\left(\frac{1}{2}\Delta K \frac{\omega^2\tau}{1+\omega^2\tau^2}-\lambda\right)\frac{\partial |X_0|^2}{\partial K_1} &=&  0, \label{eq:elK1} \\
\frac{1}{2}\Delta K |X_0|^2\frac{\partial}{\partial\tau}\left(\frac{\omega^2\tau}{1+\omega^2\tau^2}\right)
& + & \nonumber \label{eq:eltau}\\
 \left(\frac{1}{2}\Delta K \frac{\omega^2\tau}{1+\omega^2\tau^2}-\lambda\right)\frac{\partial |X_0|^2}{\partial\tau} &  =&   0, \\
|X_0|^2 &  = & X_\mathrm{L}^2.  \label{eq:elconstraint}
\end{eqnarray}
From the first of these equations we can distinguish two different cases based on which of the two factors are zero: Case A where $\partial|X_0|^2/\partial K_1 = 0$  and Case B where  $\lambda = (\Delta K/2)\omega^2\tau/(1+\omega^2\tau^2)$.

 For Case A, (\ref{eq:elK1}) reduces to (\ref{eq:stiffness}) while (\ref{eq:eltau}) determines the value of $\lambda$ which is of little concern. Therefore we are left with (\ref{eq:elconstraint}) in which we insert (\ref{eq:stiffness})  
to obtain 
\begin{equation}
(\omega\tau)^2 - \frac{M}{2F_0/F_\mathrm{c}-1}(\omega\tau) + 1 = 0
\end{equation}
or 
\begin{equation}
\omega\tau = \frac{M/2}{2F_0/F_\mathrm{c}-1}\left[1\pm\sqrt{1-\left(
\frac{2F_0/F_\mathrm{c}-1}{M/2}
\right)^2}\right].
\label{eq:loadhc2}
\end{equation}
Substituting this result back into  (\ref{eq:stiffness}), we get 
\begin{equation}
K_1 = m\omega^2 + \frac{1}{2}\Delta K\left[1\mp\sqrt{1-\left(
\frac{2F_0/F_\mathrm{c}-1}{M/2}
\right)^2}\right].
\label{eq:stiffnesshc2}
\end{equation}
We note that these solutions are only real if 
$(1-M/2)/2\le F_0/F_\mathrm{c}\le (1+M/2)/2$. The upper of the limits is equal to the critical limit (\ref{eq:fcritical2}) for displacement limitation when coupling is weak, i.e. when $M<2$. Hence these solutions are not optimal for weak coupling, but can be for high coupling. The lower limit is negative for strong coupling, so the actual lower limit will be critical force for displacement limitation $F_\mathrm{c}$ in (\ref{eq:fcritical}), i.e. the force range for these two solutions is  
\begin{equation}
1 \le \frac{F_0}{F_\mathrm{c}}\le 
\frac{1}{2}\left(1+\frac{M}{2}\right).
\end{equation}
Finally, we insert (\ref{eq:loadhc2}) and (\ref{eq:stiffnesshc2}) into the expression (\ref{eq:power}) for power and obtain 
\begin{equation}
\frac{P}{P_\mathrm{c}} = 2 \frac{F_0}{F_\mathrm{c}}-1.
\end{equation} 

Considering Case B, the second term on the l.h.s. of (\ref{eq:eltau}) is zero by assumption. For the remaining first term to be zero as required, we must have $\omega\tau=1$.  When this result is inserted into the expression (\ref{eq:X0}) for $X_0$ and (\ref{eq:elconstraint}) is used, we obtain an equation for the stiffness that can be solved to yield 
\begin{equation}
K_1 = m\omega^2 + \frac{1}{2}\Delta K\left[1\pm\frac{4}{M}\sqrt{\frac{F_0^2}{F_\mathrm{c}^2}-\frac{1}{4}\left(1+\frac{M}{2}\right)^2}\right].
\end{equation}   
Only one of these two solution branches can be taken to arbitrary large forces as the other branch goes to zero at a certain value. We note that these solutions are only valid for 
\begin{equation}
 \frac{F_0}{F_\mathrm{c}}\ge 
\frac{1}{2}\left(1+\frac{M}{2}\right).
\end{equation}
The power is again obtained by substitution into (\ref{eq:power}) yielding the constant values 
\begin{equation}
\frac{P}{P_\mathrm{c}} = \frac{M}{2}.
\end{equation}
Hence, we ultimately have saturated power for both weak and strong coupling.

\begin{figure*}[!t]
\centering
\includegraphics[]{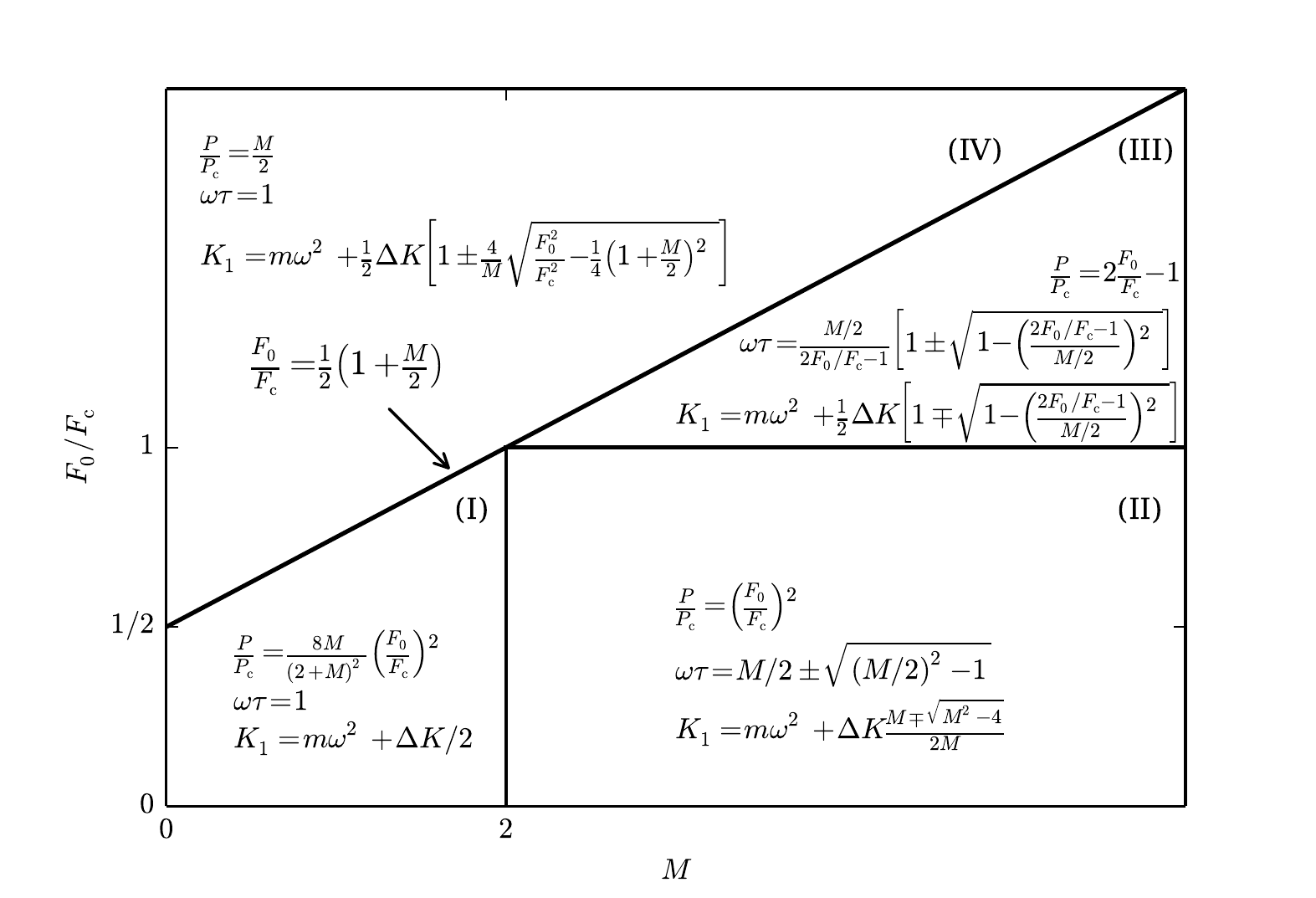}%
\caption{Operating chart of displacement limited two-port generator.}
\label{fig:opchart}
\end{figure*}
The optimal operation of the two-port harvester is summarized in Fig. \ref{fig:opchart} which divides the $M-F_0$ plane into four different domains. Below domain (IV) which has saturated power,  there is, dependent on whether $M<2$ or not, either one force range of quadratic power-dependence on forcing (I), or one range of quadratic (II) followed by linear (III) dependency on forcing. The domain-(I) and -(II) distinction is similar to the bifurcation in \cite{Renno2009}. The powers in domains (II) and (III) are identical to the VDRG power\cite{Mitcheson2004b}.

\begin{table}[!t]
\renewcommand{\arraystretch}{1.3}
\caption{Efficiency and effectiveness in each operating domain}
\label{table:eff}
\centering
\begin{tabular}{cccl}
\hline\hline
Domain & Efficiency & Effectiveness & Conditions\\
       &   $\eta$  &  $E_\mathrm{H}$ \\ 
\hline
I & $\frac{M}{M+2} $  & $\frac{4M}{(2+M)^2}\frac{F_0}{F_\mathrm{c}}$ &  
$\frac{F_0}{F_\mathrm{c}} \le \frac{1}{2}\left(1+\frac{M}{2}\right)$, $M \le 2$ 
\\
II & $\frac{1}{2} $  &  $\frac{1}{2}\frac{F_0}{F_\mathrm{c}}$ & 
$\frac{F_0}{F_\mathrm{c}} \le 1$, $M \ge 2$ 
\\
III & $1-\frac{F_\mathrm{c}}{2F_0}$  &   $1-\frac{F_\mathrm{c}}{2F_0}$ & 
$1\le \frac{F_0}{F_\mathrm{c}} \le \frac{1}{2}\left(1+\frac{M}{2}\right)$, $M \ge 2$ 
\\
IV & $\frac{M}{M+2}$ &  $\frac{M}{2}\frac{F_\mathrm{c}}{2F_0}$ &
$\frac{F_0}{F_\mathrm{c}} \ge \frac{1}{2}\left(1+\frac{M}{2}\right)$  \\
\hline\hline
\end{tabular}
\end{table}
The energy harvester effectiveness $E_\mathrm{H}$ \cite{Mitcheson2008} is a useful and increasingly used figure of merit for vibration energy harvesters.
 It is defined as the energy harvester output power divided by the maximum output power of a VDRG without parasitic losses. Hence it tells how good a harvester is compared to an ideal. It is generally different from the efficiency which is the ratio of ouput power to input power. 
In our notation here, we can write $E_\mathrm{H}=PF_\mathrm{c}/2P_\mathrm{c}F_0$. The effectiveness is listed for the different domains of optimal operation in Table \ref{table:eff}. Despite of efficiency and effectiveness being different quantities, they coincide in some of the cases due to the similarity between the two-port model and the VDRG that it is compared to. The maximum value of effectiveness is found along the border to domain (IV) and is 
\begin{equation}
\max E_\mathrm{H} = \frac{M}{M+2}
\end{equation}   
which is dependent only on the resonator figure of merit. In particular, it is independent of the displacement limit.

\section{The effect of parasitic resistances}
Parasitic conductive paths or series resistances are sometimes important, e.g, piezoelectric devices may have non-negligible leakage currents \cite{Lei2014} and electromagnetic devices can be prone to coil series resistances \cite{Maurath2012}. We account for these parasitics by considering, for the electrostatic and piezoelectric devices (Fig.\ref{fig:espe}), a total resistance $R$ on the output port that is  constituted by a load resistance $R_\mathrm{L}$  in parallel with a parasitic resistor $R_\mathrm{P}$. For the electromagnetic devices (Fig.\ref{fig:em}), we consider instead a parasitic resistance $R_\mathrm{P}$ in series with the load resistance. The overall time constant $\tau$ is then given by (\ref{eq:taues}) or (\ref{eq:tauem}) with $R$ now being the total resistance.
Parametrizing by respectively $\tau_\mathrm{P}=R_\mathrm{P}C_0$ and $\tau_\mathrm{P}=L_0/R_\mathrm{P}$ for the two cases, the output power  delivered to the load resistance can be written as
\begin{equation}
P = \frac{1}{2}\Delta K \left(1-\frac{\tau}{\tau_\mathrm{P}}\right)\frac{\omega^2\tau}{1+(\omega \tau)^2} |X_0|^2.
\label{eq:powerp}
\end{equation}
Note that $\tau \in [0,\tau_\mathrm{P}]$ and that letting $\tau_\mathrm{P}\rightarrow\infty$ recovers the case without stray parasitic resistances. 

The power (\ref{eq:powerp}) is more difficult to optimize than the previous case  (\ref{eq:power}), but for displacement-constrained operation, equations can  be found in the same way using a Lagrange multiplier. We find a parameter range of saturated power analogous to domain (IV) in Fig. \ref{fig:opchart} that has 
\begin{IEEEeqnarray}{lCl}
  \frac{P}{P_\mathrm{c}} & = & \frac{M}{2}\frac{\sqrt{(\omega\tau_\mathrm{P})^2+1}-1}{\omega\tau_\mathrm{P}},  \label{eq:powstrayfirst}\\
  \omega\tau & = & \frac{\sqrt{(\omega\tau_\mathrm{P})^2+1}-1}{\omega\tau_\mathrm{P}}, \\ 
K_1 & = & m\omega^2 + \frac{1}{2}\Delta K\left[1+ \frac{1}{\sqrt{(\omega\tau_\mathrm{P})^2+1}} \right. \nonumber \\ 
 & \pm & \left.\!\!\frac{4}{M}\sqrt{\left(\frac{F_0}{F_\mathrm{c}}\right)^2-\frac{1}{4}\left(1+\frac{M}{2}\frac{\omega\tau_\mathrm{P}}{\sqrt{(\omega\tau_\mathrm{P})^2+1}}\right)^2 } \right]\!\!\!.  
\end{IEEEeqnarray}  
This solution is valid for force amplitudes larger than 
\begin{equation}
\frac{1}{2}\left(1+\frac{M}{2}\frac{\omega\tau_\mathrm{P}}{\sqrt{(\omega\tau_\mathrm{P})^2+1}}\right) F_\mathrm{c}
\end{equation}
which reduces to (\ref{eq:fcritical2}) when $\omega\tau_\mathrm{P}\rightarrow\infty$. For smaller force amplitudes there is a range of displacement-constrained operation analogous to domain (III) in Fig. \ref{fig:opchart} that has
\begin{IEEEeqnarray}{lCl} 
 \frac{P}{P_\mathrm{c}} & = & 2 \frac{F_0}{F_\mathrm{c}}\!-\!1 - \frac{1}{\omega\tau_\mathrm{P}}\!\left[\frac{M}{2} - \sqrt{\frac{M^2}{4}-\left(\!2\frac{F_0}{F_\mathrm{c}}\!-\!1\!\right)^2}\right], \\
 \omega\tau & = & \left[\frac{M}{2} - \sqrt{\frac{M^2}{4}-\left(\!2\frac{F_0}{F_\mathrm{c}}\!-\!1\!\right)^2}\right]\Bigg/\left(\!2\frac{F_0}{F_\mathrm{c}}\!-\!1\!\right), \\
K_1 & = & m\omega^2 + \frac{\Delta K}{M}\left[\frac{M}{2} - \sqrt{\frac{M^2}{4}-\left(\!2\frac{F_0}{F_\mathrm{c}}\!-\!1\!\right)^2}\right].
\label{eq:powstraylast}
\end{IEEEeqnarray} 

\begin{figure}[!t]
\centering
\includegraphics[]{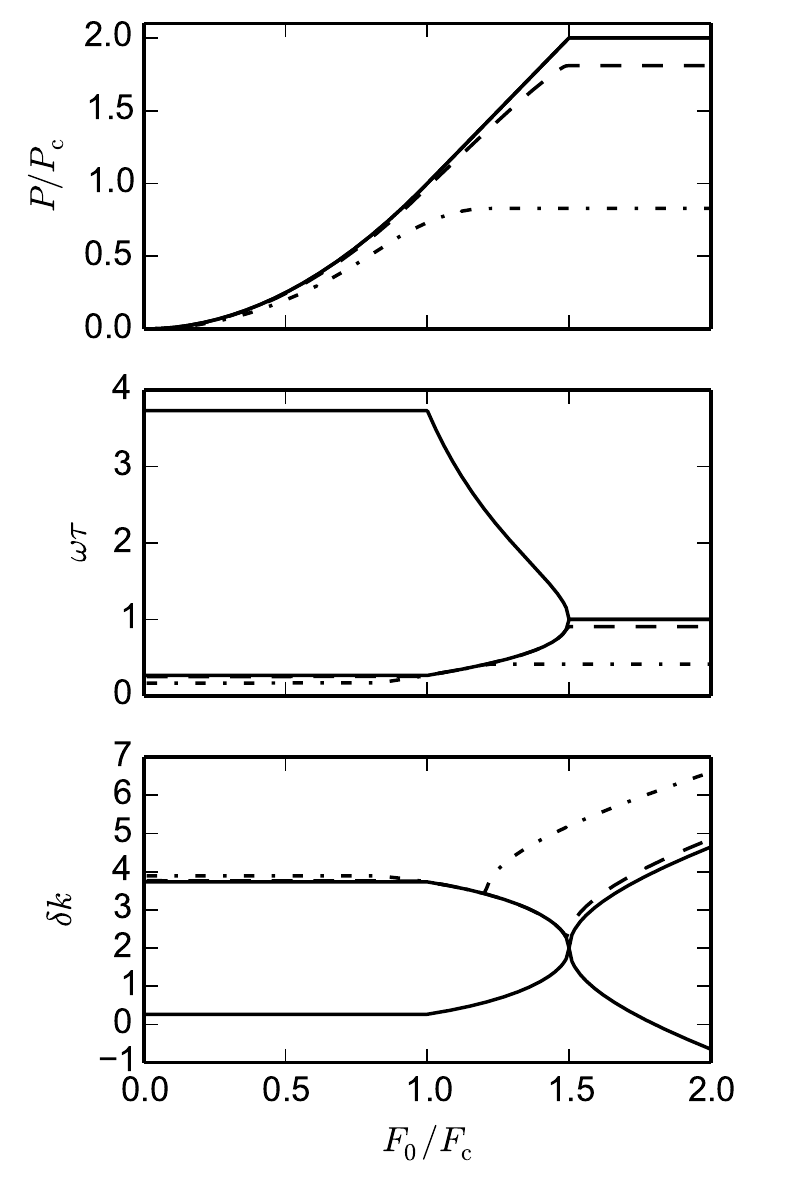}%
\caption{Optimal solution for $M=4.0$ and for $\omega\tau_\mathrm{p}=\infty$ (solid lines), $\omega\tau_\mathrm{p}=10.0$ (dashed lines) and $\omega\tau_\mathrm{p}=1.0$ (dash-dotted lines)}
\label{fig:optpow}
\end{figure}
Fig. \ref{fig:optpow} shows high-coupling examples of optimized solutions. The solutions with parasitics (finite $\omega\tau_\mathrm{P}$) are obtained by combining (\ref{eq:powstrayfirst}-\ref{eq:powstraylast}) with numerical optimization of (\ref{eq:powerp}). The numerical optimization consists in inserting (\ref{eq:stiffness}), which is still valid with the modified definition of $\tau$, into (\ref{eq:powerp}) and maximizing the resulting function  with respect to $\tau$ for the cases with $|X_0|\le X_\mathrm{L}$.  The maximization is acccomplished using optimization methods in the \textit{SciPy} module \cite{Jones2001} of the \textit{Python} programming language. The optimal stiffness is given in terms of $\delta k = (K_1-m\omega^2)/\omega b$. The parasitic resistances influence the optimum for all ranges of acceleration, but have most severe effects under displacement-limited operation. Note that the  two equivalent optima for $\omega\tau_\mathrm{P}=\infty$ reduce to one for a finite $\omega\tau_\mathrm{P}$ and that this solution is the one with the smallest time-constant, which is also seen in experiments \cite{Lei2014}.

\section{Concluding remarks}
We investigated the optimal choice of stiffness and load resistance for two-port energy harvesters driven by harmonic vibrations. Even without displacement constraints in effect, the results differ from previous works because stiffness was varied rather than the drive frequency. A complete optimization encompassing the displacement-limited operating regime was made analytically and an operating chart established. The normalized output power as a function of normalized force has only one parameter, the resonator figure of merit $M$, which also determines the maximum energy harvester effectiveness. If displacement-limits are in effect, the figure of merit should be as large as possible in order to maximize power, otherwise it need only exceed a threshold value of $M\ge 2$. The analytical expressions obtained for the effects of stray conductive paths or series resistances makes quantification of these performance degrading factors simple.  
The results of this paper can be used to judge the merit of energy harvesters with a linear two-port architecture and act as design guidance when using electric control of load and stiffness to design energy harvesting systems that operate effectively.



\begin{thebibliography}{10}
\providecommand{\url}[1]{#1}
\csname url@samestyle\endcsname
\providecommand{\newblock}{\relax}
\providecommand{\bibinfo}[2]{#2}
\providecommand{\BIBentrySTDinterwordspacing}{\spaceskip=0pt\relax}
\providecommand{\BIBentryALTinterwordstretchfactor}{4}
\providecommand{\BIBentryALTinterwordspacing}{\spaceskip=\fontdimen2\font plus
\BIBentryALTinterwordstretchfactor\fontdimen3\font minus
  \fontdimen4\font\relax}
\providecommand{\BIBforeignlanguage}[2]{{%
\expandafter\ifx\csname l@#1\endcsname\relax
\typeout{** WARNING: IEEEtran.bst: No hyphenation pattern has been}%
\typeout{** loaded for the language `#1'. Using the pattern for}%
\typeout{** the default language instead.}%
\else
\language=\csname l@#1\endcsname
\fi
#2}}
\providecommand{\BIBdecl}{\relax}
\BIBdecl

\bibitem{Mitcheson2008}
P.~D. Mitcheson, E.~M. Yeatman, G.~K. Rao, A.~S. Holmes, and T.~C. Green,
  ``Energy harvesting from human and machine motion for wireless electronic
  devices,'' \emph{Proc. {IEEE}}, vol.~96, no.~9, pp. 1457--1486, Sep. 2008.

\bibitem{Guyomar2005}
D.~Guyomar, A.~Badel, E.~Lefeuvre, and C.~Richard, ``Toward energy harvesting
  using active materials and conversion improvement by nonlinear processing,''
  \emph{{IEEE} Trans. Ultrason., Ferroelectr., Freq. Control}, vol.~52, no.~4,
  pp. 584--595, Apr. 2005.

\bibitem{Mitcheson2004b}
P.~D. Mitcheson, T.~C. Green, E.~M. Yeatman, and A.~S. Holmes, ``Architectures
  for vibration-driven micropower generators,'' \emph{J. Microelectromech.
  Syst.}, vol.~13, no.~3, pp. 429--440, Jun. 2004.

\bibitem{Eichhorn2011}
\BIBentryALTinterwordspacing
C.~Eichhorn, R.~Tchagsim, N.~Wilhelm, and P.~Woias, ``A smart and
  self-sufficient frequency tunable vibration energy harvester,'' \emph{J.
  Micromech. Microeng.}, vol.~21, no.~10, p. 104003, 2011.
\BIBentrySTDinterwordspacing

\bibitem{Sterken2003b}
T.~Sterken, P.~Fiorini, K.~Baert, R.~Puers, and G.~Borghs, ``An electret-based
  electrostatic $\mu$-generator,'' in \emph{12th International Conference on Solid-State Sensors, Actuators and Microsystems, TRANSDUCERS 2003},
  vol.~2, 2003, pp. 1291--1294 vol.2.

\bibitem{Tvedt2010}
L.~G.~W. Tvedt, D.~S. Nguyen, and E.~Halvorsen, ``Nonlinear behavior of an
  electrostatic energy harvester under wide- and narrowband excitation,''
  \emph{J. Microelectromech. Syst.}, vol.~19, no.~2, pp. 305 --316, april 2010.

\bibitem{Roundy2004}
\BIBentryALTinterwordspacing
S.~Roundy and P.~K. Wright, ``A piezoelectric vibration based generator for
  wireless electronics,'' \emph{Smart Mater. Struct.}, vol.~13, no.~5, pp.
  1131--1142, 2004.
\BIBentrySTDinterwordspacing

\bibitem{Goldschmidtboeing2011}
\BIBentryALTinterwordspacing
F.~Goldschmidtboeing, M.~Wischke, C.~Eichhorn, and P.~Woias, ``Parameter
  identification for resonant piezoelectric energy harvesters in the low- and
  high-coupling regimes,'' \emph{J. Micromech. Microeng.}, vol.~21, no.~4, p.
  045006, 2011.
\BIBentrySTDinterwordspacing

\bibitem{Amirtharajah1998}
R.~Amirtharajah and A.~P. Chandrakasan, ``Self-powered signal processing using
  vibration-based power generation,'' \emph{{IEEE} J. Solid-State Circuits},
  vol.~33, no.~5, pp. 687--695, May 1998.

\bibitem{Maurath2012}
D.~Maurath, P.~Becker, D.~Spreemann, and Y.~Manoli, ``Efficient energy
  harvesting with electromagnetic energy transducers using active low-voltage
  rectification and maximum power point tracking,'' \emph{IEEE Journal of Solid-State Circuits}, vol.~47, no.~6, pp. 1369--1380, June 2012.

\bibitem{Lefeuvre2005}
\BIBentryALTinterwordspacing
E.~Lefeuvre, A.~Badel, C.~Richard, L.~Petit, and D.~Guyomar, ``A comparison
  between several vibration-powered piezoelectric generators for standalone
  systems,'' \emph{Sens. Actuators A}, vol. 126, no.~2, pp. 405--416, Feb.
  2006. 
\BIBentrySTDinterwordspacing

\bibitem{Dicken2012}
J.~Dicken, P.~Mitcheson, I.~Stoianov, and E.~Yeatman, ``Power-extraction
  circuits for piezoelectric energy harvesters in miniature and low-power
  applications,'' \emph{{IEEE} Trans. Power Electron.}, vol.~27,
  no.~11, pp. 4514--4529, Nov 2012.

\bibitem{Miller2012}
L.~M. Miller, P.~D. Mitcheson, E.~Halvorsen, and P.~K. Wright, ``Coulomb-damped
  resonant generators using piezoelectric transduction,'' \emph{Appl. Phys.
  Lett.}, vol. 100, no.~23, pp. 233\,901 --233\,901--4, jun 2012.

\bibitem{Dhulst2006}
R.~D'hulst, P.~D. Mitcheson, and J.~Driesen, ``Cmos buck-boost power processing
  circuitry for powermems generators,'' in \emph{Power{MEMS} 2006 Technical
  Digest}, Berkeley, 2006, pp. 215--218.

\bibitem{Lefeuvre2007}
E.~Lefeuvre, D.~Audigier, and D.~Richard, C.and~Guyomar, ``Buck-boost converter
  for sensorless power optimization of piezoelectric energy harvester,''
  \emph{{IEEE} Trans. Power Electron.}, vol.~22, no.~5, pp. 2018--2025, 2007.

\bibitem{Dhulst2010}
R.~D'hulst, T.~Sterken, R.~Puers, G.~Deconinck, and J.~Driesen, ``Power
  processing circuits for piezoelectric vibration-based energy harvesters,''
  \emph{{IEEE} Transactions on Industrial Electronics}, vol.~57, no.~12, pp.
  4170 --4177, dec. 2010.

\bibitem{duToit2005}
N.~E. duToit, B.~L. Wardle, and S.~G. Kim, ``Design considerations for
  mems-scale piezoelectric mechanical vibration energy harvesters,''
  \emph{Integr. Ferroelectr.}, vol.~71, pp. 121--160, Jul. 2005.

\bibitem{Renno2009}
\BIBentryALTinterwordspacing
J.~M. Renno, M.~F. Daqaq, and D.~J. Inman, ``On the optimal energy harvesting
  from a vibration source,'' \emph{J. Sound Vibr.}, vol. 320, no. 1–2, pp. 386
  -- 405, 2009.
\BIBentrySTDinterwordspacing

\bibitem{Williams1996}
C.~B. Williams and R.~B. Yates, ``Analysis of a micro-electric generator for
  microsystems,'' \emph{Sens. Actuators A}, vol.~52, no. 1-3, pp. 8--11, 1996.

\bibitem{Halvorsen2013c}
\BIBentryALTinterwordspacing
E.~Halvorsen, C.~P. Le, P.~D. Mitcheson, and E.~M. Yeatman,
  ``Architecture-independent power bound for vibration energy harvesters,''
  \emph{Journal of Physics: Conference Series}, vol. 476, no.~1, p. 012026,
  2013.
\BIBentrySTDinterwordspacing

\bibitem{Renaud2012}
\BIBentryALTinterwordspacing
M.~Renaud, R.~Elfrink, M.~Jambunathan, C.~de~Nooijer, Z.~Wang, M.~Rovers,
  R.~Vullers, and R.~van Schaijk, ``Optimum power and efficiency of
  piezoelectric vibration energy harvesters with sinusoidal and random
  vibrations,'' \emph{J. Micromech. Microeng.}, vol.~22, no.~10, p. 105030,
  2012.
\BIBentrySTDinterwordspacing

\bibitem{Tilmans1996}
H.~A.~C. Tilmans, ``Equivalent circuit representation of electromechanical
  transducers: I. lumped-parameter systems,'' \emph{J. Micromech. Microeng.},
  vol.~6, pp. 157--176, 1996.

\bibitem{Tilmans1996errata}
------, ``Equivalent circuit representation of electromechanical transducers
  .1. lumped-parameter systems (vol 6, pg 157, 1996),'' \emph{J. Micromech. Microeng.}, vol.~6, no.~3, pp. 359--359, Sep. 1996.

\bibitem{Ikeda1996}
T.~Ikeda, \emph{Fundamentals of Piezoelectricity}.\hskip 1em plus 0.5em minus
  0.4em\relax New York: Oxford University Press Inc., 1996.

\bibitem{Senturia2001}
S.~Senturia, \emph{Microsystem Design}.\hskip 1em plus 0.5em minus 0.4em\relax
  Kluwer Academic Publishers, 2001.

\bibitem{Tadmor2003}
E.~B. Tadmor and G.~K{\' o}sa, ``Electromechanical coupling correction for a
  piezoelectric layered beam,'' \emph{Journal of microelectromechanical
  systems}, vol.~12, pp. 899--906, 2003.

\bibitem{Vittoz2010}
\BIBentryALTinterwordspacing
E.~Vittoz, ``\BIBforeignlanguage{English}{Quartz and mem resonators},'' in
  \emph{\BIBforeignlanguage{English}{Low-Power Crystal and MEMS Oscillators}},
  ser. Integrated Circuits and Systems.\hskip 1em plus 0.5em minus 0.4em\relax
  Springer Netherlands, 2010, vol.~0, pp. 7--22.
\BIBentrySTDinterwordspacing

\bibitem{Lei2014}
A.~Lei, R.~Xu, L.~Borregaard, M.~Guizzetti, O.~Hansen, and E.~Thomsen,
  ``Impedance based characterization of a high-coupled screen printed pzt thick
  film unimorph energy harvester,'' \emph{Journal of Microelectromechanical Systems }, vol.~23, no.~4, pp. 842--854, Aug 2014.

\bibitem{Jones2001}
\BIBentryALTinterwordspacing
E.~Jones, T.~Oliphant, P.~Peterson \emph{et~al.}, ``{SciPy}: Open source
  scientific tools for {Python},'' 2001--, [Online; accessed 2016-02-22].
  Available: \url{http://www.scipy.org/}
\BIBentrySTDinterwordspacing

\end{thebibliography}
\end{document}